\documentclass [prl,twocolumn,superscriptaddress,floatfix,amsmath,amssymb] {revtex4-2}
\usepackage{amsmath}
\usepackage{graphicx}
\usepackage{hyperref}
\usepackage{color}
\usepackage{amssymb}
\usepackage{bm}
\usepackage{babel}

\newcommand{\beq} {\begin{equation}}
\newcommand{\eeq} {\end{equation}}
\newcommand{\bea} {\begin{eqnarray}}
\newcommand{\eea} {\end{eqnarray}}
\newcommand{\be} {\begin{equation}}
\newcommand{\ee} {\end{equation}}

\renewcommand{\)}{\right)}

\definecolor{darkgreen}{RGB}{0,170,0}

\begin{document}
\title{Density of states and spectral function of a superconductor out of a quantum-critical metal}
\author{Shang-Shun Zhang}
\affiliation{School of Physics and Astronomy and William I. Fine Theoretical Physics Institute,
University of Minnesota, Minneapolis, MN 55455, USA}
\author{Andrey V. Chubukov}
\affiliation{School of Physics and Astronomy and William I. Fine Theoretical Physics Institute,
University of Minnesota, Minneapolis, MN 55455, USA}
\date{\today}
\begin{abstract}
We analyze the validity of a quasiparticle description of a superconducting state at a metallic quantum-critical point (QCP).
A normal state at a QCP is a non-Fermi liquid with no coherent quasiparticles.
A superconducting order gaps out low-energy excitations, except for a sliver of states for non-s-wave gap symmetry, and
at a first glance, should restore a coherent quasiparticle behavior.
We argue that this does not necessarily hold as in some cases the fermionic self-energy remains singular slightly above the gap edge.
This singularity gives rise to markedly non-BCS behavior of the density of states and to broadening and eventual vanishing of the quasiparticle peak in the spectral function.
We analyze the set of quantum-critical models with an effective dynamical 4-fermion interaction, mediated by a gapless boson at a QCP, $V(\Omega) \propto 1/\Omega^\gamma$.
We show that coherent quasiparticle behavior in a superconducting state holds for $\gamma <1/2$, but breaks down for larger $\gamma$.
We discuss signatures of quasiparticle breakdown and compare our results with the data.
\end{abstract}
\maketitle

{\it {\bf Introduction.}}~~~Metals near a  quantum critical point (QCP) display a number of non-Fermi liquid properties
like linear-in-$T$ resistivity,
a broad peak in the spectral function near $k_F$ with linear-in-$\omega$ width,
singular behavior of optical conductivity, etc~\cite{martin1990normal,lohneysen1994non,Norman1998,shen1999novel,
grigera2001magnetic,Damascelli2003,Inosov2007,daou2009linear,cooper2009anomalous,sarkar2017fermi,*jin2011link,
hashimoto2014energy,Kaminski2015,keimer2015quantum,taillefer_annual,legros2019universal,varma,Hartnoll2022,Dessau2022}.
 These properties are often thought to be caused by the coupling of fermions to near-gapless fluctuations of an order parameter, which condenses at a QCP~\cite{peter_review,valla1999,acs,*finger_2001,Fink2006,shibauchi2014quantum,restrepo2022,Scalapino2012,efetov,review2,max_2,
torroba_2}.
The same fermion-boson interaction gives rise to superconductivity near a QCP~\cite{nick_b,acf,Chubukov_2020a,max_last,ital2,Subir2,wang_13,Wang2016,Wang_H_17,son,*son2,
paper_1,yuz_3,zhang_22}.

A superconducting order gaps out low-energy excitations, leaving at most a tiny subset of gapless states for a non-$s-$wave order parameter.
A general belief
 has been
that this restores fermionic coherence.
A frequently cited experimental evidence is the observed re-emergence of a quasiparticle peak below $T_c$ in near-optimally doped cuprates (see e.g., Ref.~\cite{Kaminski2000}).
From theory side, the argument is that the fermionic self-energy in a superconductor has a conventional Fermi-liquid form $\Sigma(\omega) \sim \omega$ at the lowest $\omega$,
in distinction from a non-Fermi-liquid $\Sigma (\omega) \propto \omega^a$ with $a <1$ in the normal state
~\cite{Oganesyan2001,Metzner2003,*DellAnna2006,rech_2006,metlitski2010quantum1,*metlitski2010quantum2,senthil,
raghu_15,*Fitzpatrick_15,avi,sslee,*sslee2,*lunts_2017,punk,cm_pom}.
In this paper, we analyze theoretically whether fermions in a superconducting state at a QCP can be viewed as well-defined coherent quasiparticles.
We argue that this is not necessarily the case as fermionic self-energy can still be singular on a real frequency axis immediately above the gap edge.
This singularity gives rise to markedly non-BCS behavior of the density of states (DoS) and to broadening and eventual vanishing of the quasiparticle peak.

\begin{figure}
\centering
\includegraphics[scale=1]{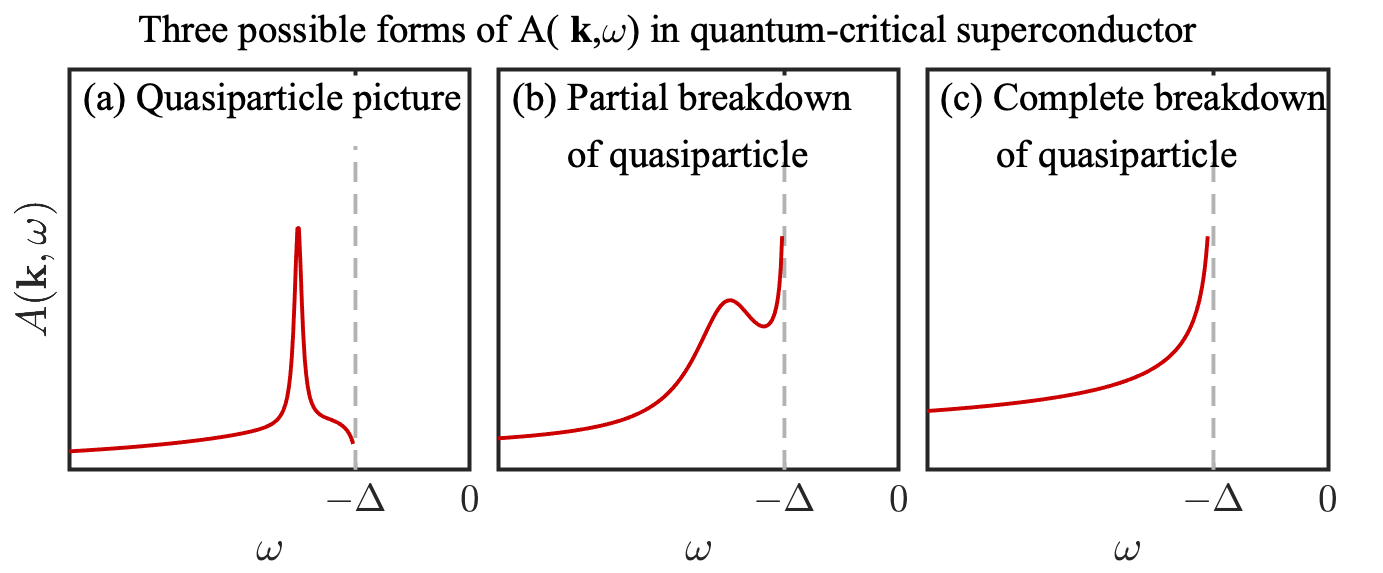}\caption{Three possible forms of the electronic spectral function $A({\bm k}, \omega)$ at $T=0$ in a quantum critical superconductor at a small but finite $k-k_F$  and in the absence of impurity broadening.
(a): $A({\bm k}, \omega)$   vanishes at
$\rvert \omega \rvert = \Delta$
and  has a well-defined peak at $\omega > \Delta$,  (b): $A({\bm k}, \omega)$ diverges at
$\rvert \omega \rvert = \Delta$,
but it non-monotonic at larger $\omega$. The
peak in $A({\bm k}, \omega)$ at
$\rvert \omega \rvert > \Delta$
broadens, but still exists. (c): $A({\bm k}, \omega)$ diverges at
$\rvert \omega \rvert = \Delta$,
and monotonically decreases at larger $\omega$.
   In case (a) fermions
   can be viewed as
   well-defined quasiparticles, in
   case (c)  the quasiparticle picture completely breaks down. The  case (b) is the intermediate one
   between (a) and (c).
 }
\label{fig:sche}
\end{figure}

\begin{figure*}
\centering
\includegraphics[scale=1]{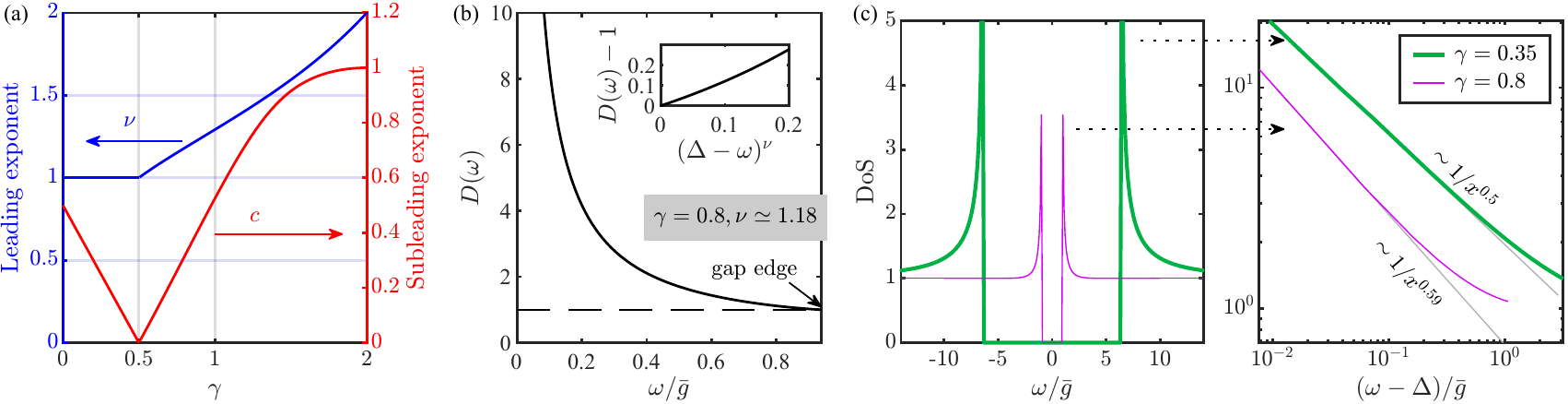}\caption{
(a) Exponents $\nu$ and $c$ for the leading and the subleading terms in  the expansion $D(\omega) \simeq 1+\alpha (\Delta-\omega)^\nu + \beta (\Delta - \omega)^{\nu + c}$, where $D(\omega) = \Delta (\omega)/\omega$ and
 the gap edge $\Delta$ is the solution of $D (\omega = \Delta) =1$.
(b) Numerical result for $D(\omega)$ for $\gamma=0.8$. Inset shows the
power-law behavior near the gap edge with $\nu = 1.18$, consistent with (a).
(c) Fermionic DoS at $T=0$ for $\gamma=0.35$ (thick green line) and $\gamma=0.8$ (thin pink line).
In both cases, the DoS vanishes below the gap edge $\Delta$ and
has a power-law  singularity above it
$N(\omega) \propto 1/(\omega -\Delta)^{\nu/2}$,
but the exponent
$\nu$ is different in the two cases,
as we show in
the right panel.
}
\label{fig:exponent}
\end{figure*}

For superconductivity away from a QCP,
mediated
by a massive boson, numerous earlier studies have found that
the spectral function
$A({\bm k}, \omega)$
at $T=0$ has a $\delta$-functional peak at
$\omega = (\Delta^2 + (\xi_{\bm k}/Z)^2)^{1/2}$,
where
 ${\xi_{\bm k}} = v_F (k-k_F)$
 is a fermionic dispersion ($v_F$ is a Fermi velocity), $\Delta$ is a superconducting gap, and $Z$ is an inverse quasiparticle residue.
A
$\delta$-functional peak  holds for momenta near the Fermi surface, as long as $\omega  < \Delta + \omega_0$, where $\omega_0$ is a mass of a pairing boson in energy units.
At larger $\omega$, fermionic damping kicks in, and the peak broadens.
The same physics leads to peak-dip-hump behavior of
$A({\bm k}, \omega)$
as a function of $\omega$, observed most spectacularly in near-optimally doped
 cuprate Bi$_2$Sr$_2$CaCu$_2$O$_{8+\delta}$ (see, e.g,
Refs.~\cite{Dessau1991,Kordyuk_pdh}).
 At a QCP, the pairing boson becomes massless and $\omega_0$ vanishes.
 This creates a singular behavior near the gap edge at $\omega =\Delta$, which holds even when
${\xi_{\bm k}}$ is finite.

A simple experimentation shows that there are three possible forms of $A({\bm k}, \omega)$, which we present in Fig.~\ref{fig:sche}:
it (i) either vanishes at $\omega = \Delta$ and has a well-defined peak at $\omega > \Delta$ whose width at small ${\xi_{\bm k}}$ is
parametrically smaller than its energy;
or (ii) diverges at $\omega = \Delta$, but is non-monotonic at larger $\omega$ and displays a broad maximum at some $\omega > \Delta$,
or (iii)  diverges at $\omega = \Delta$ and monotonically decreases at larger $\omega$.
In the first case, fermions in a quantum-critical superconductor can be viewed as well-defined quasiparticles;
in the last case the quasiparticle picture completely breaks down;
the second case is the intermediate one between the other two.
Our goal is to understand under what circumstances $A({\bm k}, \omega)$ of a quantum-critical superconductor has one of these forms.

{\it {\bf Model.}}~~~ For our study, we consider dispersion-full fermions, Yukawa-coupled to a massless boson.
We assume, like in earlier works~(see, e.g., Refs.~\cite{zhang2023free}),
that a boson is Landau overdamped, and its effective velocity is far smaller than $v_F$.
In this situation, the interaction that gives rise to non-Fermi liquid in the normal state and to superconductivity, is a purely dynamical $V(\Omega)$.
The fermionic self-energy and the pairing gap, tuned into a proper spatial pairing channel, are then determined by two coupled equations
in the frequency domain.
At a QCP, $V(\Omega)$  is singular at vanishing $\Omega$ in spatial dimension $D \leq 3$, and behaves as $V(\Omega) \propto ({\bar g}/\Omega)^\gamma$,
where ${\bar g}$ is the effective fermion-boson coupling, and the exponent $\gamma$ is determined by the underlying microscopic model.
The most studied models of this kind are of fermions near an Ising-nematic or Ising/ferromagnetic QCP ($\gamma =1/3$) and near an antiferromagnetic or charge density wave QCP $(\gamma =1/2)$.
The same effective interaction emerges for dispersion-less fermions in a quantum dot
coupled to Einstein bosons (the Yuakawa-SYK model)~\cite{Schmalian_19,*Schmalian_19a,Wang_19,Chowdhury_2020,Classen21}.
For this last case, the exponent $\gamma$ is a continuous variable $\gamma \in (0,1)$, depending on the ratio of fermion and boson flavors.
An extension of the Yukawa-SYK model to $\gamma \in (1,2)$ has recently been proposed~\cite{joerg}.
We follow these works and consider $\gamma$ as a continuous variable.
We note that the value of $\gamma$ is generally larger deep in a superconducting state because of feedback from superconductivity on the bosonic polarization.
For simplicity, we neglect potential in-gap states associated with
non-$s$-wave pairing symmetry and focus on the spectral function of fermions away from the nodal points and on features in the density of states (DoS) above the gap edge.
An extension to models with in-gap states is straightforward.

In previous studies of the $\gamma$-model, we focused on the novel superconducting behavior at $\gamma >1$,
when the pairing interaction is attractive on the Matsubara axis, while on the real axis Re$V(\Omega)$ is repulsive~\cite{paper_4,paper_5}.
We argued that this dichotomy gives rise to phase slips of the gap function on the real axis.
Here, we restrict ourselves to $\gamma \leq 1$, when this physics is not present and, hence, does not interfere with the analysis of the validity of a quasiparticle description in a superconducting state.\\

{\it {\bf Pairing gap and  quasiparticle residue.}}~~~
For superconductivity mediated by a dynamical interaction, the paring gap  $\Delta (\omega)$ and the inverse quasiparticle residue
$Z (\omega)$ are functions of the running real fermionic frequency $\omega$. We define the gap edge $\Delta$ (often called the gap)
from the condition $\Delta (\omega) = \omega$ at $\omega = \Delta$.

For our purposes, it is convenient to introduce $D (\omega) = \Delta (\omega)/\omega$. The gap edge is at $|D|=1$.
The equation for $D(\omega)$ that we need to solve is
\bea
\omega B(\omega)D(\omega)=A(\omega)+C(\omega),\label{eq:gap_real1}
\eea
where $B(\omega)$ and $A(\omega)$ are regular functions of $\omega$ (see ~\cite{SM,Marsiglio_88}).
The $C(\omega)$ term depends on the running $D(\omega)$,
\begin{equation}
C(\omega)=\bar{g}^{\gamma}\sin\frac{\pi\gamma}{2}\int_{0}^{\omega}\frac{d\Omega}{\Omega^{\gamma}}
\frac{D(\omega-\Omega)-D(\omega)}{\sqrt{D^{2}(\omega-\Omega)-1}}.\label{eq:Cw}
\end{equation}
Its presence makes Eq.~(\ref{eq:gap_real1}) an integral equation.
The inverse residue $Z(\omega)$ is expressed via  $D(\omega')$ as
\bea
Z(\omega)= B(\omega) + \frac{\bar{g}^{\gamma}\sin\frac{\pi\gamma}{2}}{\omega}\int_{0}^{\omega}\frac{d\Omega}{\Omega^{\gamma}}
\frac{1}{\sqrt{D^{2}(\omega-\Omega)-1}} \label{eq:gap_real2}
\eea
 and is readily obtained once $D(\omega)$ is known.
\begin{figure*}
\centering
\includegraphics[scale=0.9]{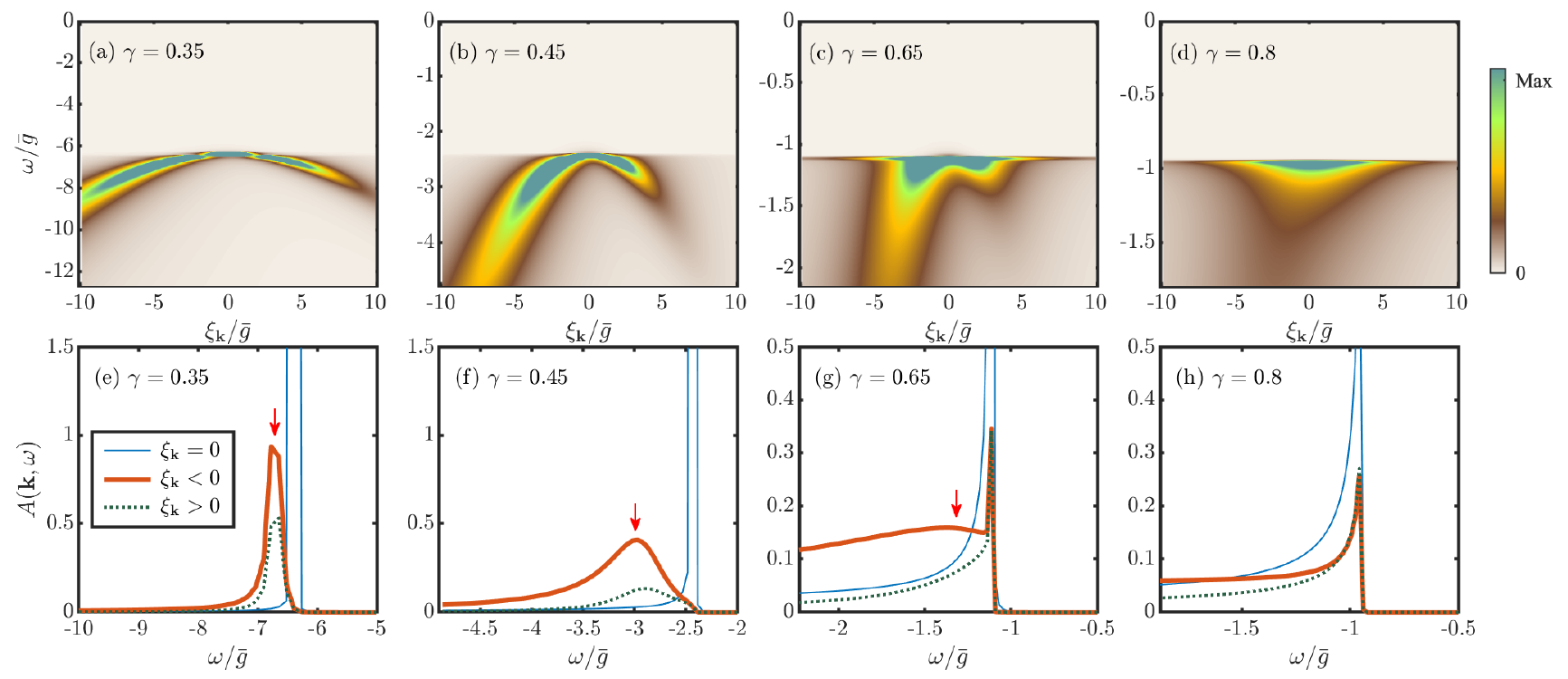}\caption{
Spectral function $A({\bm k},\omega)$ at $T=0$ for four representative $\gamma$.
The broadening in the plots is intrinsic.
(a-d): color-coded plot at negative $\omega$, as measured by the ARPES intensity at T = 0.
(e-f): constant-${\bm k}$ cuts of $A({\bm k},\omega)$ at $\xi_{\bm k} =0$ and at $\xi_{\bm k} = \pm 4 \bar{g}$.
For $\gamma <1/2$, the spectral function has a sharp quasiparticle peak at $\omega +\Delta \propto \xi^2_{{\bm k}}$.
For $\gamma >1/2$, the peak moves to $\omega +\Delta \propto |\xi_{{\bm k}}|^{1/(1-\gamma)}$ and broadens up, which eventually disappears (see text).
}
\label{fig:spectral_sc}
\end{figure*}

At $\gamma =0$, which models a BCS superconductor, $C(\omega) =0$ and $D(\omega) = A(\omega)/(\omega B(\omega))$ is
a regular function of frequency. Near the gap edge at $\omega >0$, $D(\omega) -1 \sim \omega -\Delta$ and $Z(\omega) \approx Z(\Delta) \equiv Z$.
We assume and then verify that $D(\omega)$ remains regular in some range of $\gamma >0$.
Substituting $D(\omega) -1 \sim \omega -\Delta$ into (\ref{eq:Cw}) for $\gamma >0$, we obtain $C(\omega) - C(\Delta) \sim (\omega -\Delta)^{3/2 -\gamma}$.
We see that $C(\omega)$ is non-analytic near the gap edge, but for $\gamma <1/2$, the exponent $3/2-\gamma$ is larger than one.
In this situation, the non-analytic term in $C(\omega)$ generates a non-analytic term in $D(\omega)$ of order $(\omega -\Delta)^{3/2 -\gamma}$,
which is smaller than the regular $\omega -\Delta$  term.
Evaluating the prefactors,  we obtain slightly above the gap edge, at $\omega = \Delta + \delta$
\bea\label{eq:gap1b}
&&D^{\prime}(\Delta+\delta) = 1 + \alpha \delta + A \cos[\pi (3/2-\gamma)] \delta^{3/2-\gamma}, \nonumber \\
&&
D^{\prime\prime}(\Delta +\delta) = - A \sin[\pi (3/2-\gamma)]\delta^{3/2-\gamma},
\eea
where
$\alpha \sim 1/{\bar g}$, $A = \sqrt{ {\alpha \over 2} } { {\bar g}^{\gamma} \sin(\pi \gamma/2) \over \Delta B(\Delta) } J (\gamma,1)$ and
 $J(\gamma,\nu)$ is expressed via Beta functions:
\begin{equation}
J(\gamma,\nu)=B(1-\gamma,\gamma-1-{\nu\over2})-B(1-\gamma,\gamma-1+{\nu\over2}).
\end{equation}
For $\gamma >1/2$, $3/2 -\gamma >1$, and the calculation of $D(\omega)$ has to be done differently.
We find after straightforward analysis that the leading $\delta$-dependent term in $D(\Delta + \delta)$ is non-analytic and of order $\delta^\nu$, where $\nu$ is the solution of $J(\gamma,\nu) =0$. The exponent $\nu \approx 1+0.67 (\gamma -1/2)$ for $\gamma \approx 1/2$ and $\nu  \approx 1.3$ for $\gamma =1$.
The subleading term in $D(\Delta + \delta)$ scales as $\delta^{\nu +c}$, where $c >0$ is
approximately
linear in $\gamma -1/2$.
In Fig.~\ref{fig:exponent}, we plot $\nu (\gamma)$ and $c (\gamma)$ along with the numerical results of $D(\omega)$ for a representative $\gamma = 0.8$.
The exponent $\nu$ extracted from this numerical $D(\omega)$ is $1.18$, which matches perfectly with the analytical result.
The behavior at $\gamma =1/2$ is special, and we discuss it in Ref.~\cite{SM}.

Substituting $D(\Delta + \delta)$ into the formula for $Z(\omega)$, Eq. (\ref{eq:gap_real2}), we obtain
 \bea
Z^{\prime}(\Delta + \delta) \!\! &=&\!\! Z (\Delta) \! +\! B \cos(\pi(\gamma+\nu/2-1)) \delta^{1-\gamma-\nu/2}, \label{eq:Zw_2a} \\
Z^{\prime \prime}(\Delta + \delta) \!\! &=&\!\!  B \sin(\pi(\gamma+\nu/2-1)) \delta^{1-\gamma-\nu/2}.  \label{eq:Zw_2b}
\eea
where $B = \frac{\bar{g}^{\gamma}\sin\frac{\pi\gamma}{2}}{\Delta\sqrt{2\alpha}} B(1-\gamma,\frac{\nu}{2}+\gamma-1)$.
For $\gamma <1/2$,
$Z (\omega) = Z(\Delta) + O(\delta^{1/2 -\gamma})$
is approximately a constant near the gap edge. For $\gamma >1/2$, the inverse residue diverges at the gap edge, indicating a qualitative change in the system behavior.\\

{\it {\bf Spectral function and DoS.}}~~
The spectral function and the  DoS per unit volume are given by
\bea
&&A({\bm k}, \omega) = -\frac{1}{\pi} {\text {Im}} G_R ({\bm k}, \omega), \nonumber \\
&& N (\omega) = \frac{1}{V} \sum_{\bm k} A({\bm k}, \omega) = N_F \omega {\text {Im}}
\sqrt{\frac{1}{\Delta^2 (\omega) -\omega^2}},
\label{A}
\eea
where the retarded Green's function $G_R ({\bm k}, \omega) = -(\omega Z(\omega) + {\xi_{\bm k}})/({\xi_{\bm k}^2} + (\Delta^2 (\omega)-\omega^2) Z^2(\omega))$.
ARPES intensity is proportional to $A({\bm k}, \omega) n_F (\omega)$, which at $T =0$ selects negative $\omega$.
At $\gamma =0$ (BCS limit), $N(\omega) \sim 1/(\omega -\Delta)^{1/2}$, and the spectral function has a $\delta$-functional
peak at  $\omega = (\Delta^2 + ({\xi_{\bm k}}/Z)^2)^{1/2}$. In Fig.~\ref{fig:exponent} (c,d), we show the DoS $N(\omega)$,
obtained from the numerical solution of the full gap equation (\ref{eq:gap_real1}) for representative $\gamma =0.35$ and $0.8$.
We see that in both cases the DoS describes a gapped continuum, but there is a qualitative difference in the behavior near the gap edge:
for $\gamma =0.35$, $ N(\omega)$  has the same $1/\delta^{1/2}$ singularity as for $\gamma =0$, and for $\gamma =0.8$ the DOS
behaves as $1/\delta^{0.59}$, which perfectly matches the analytical form $\delta^{-\nu/2}$, given that $\nu =1.18$ for $\gamma =0.8$.

\begin{figure}
\centering
\includegraphics[scale=0.9]{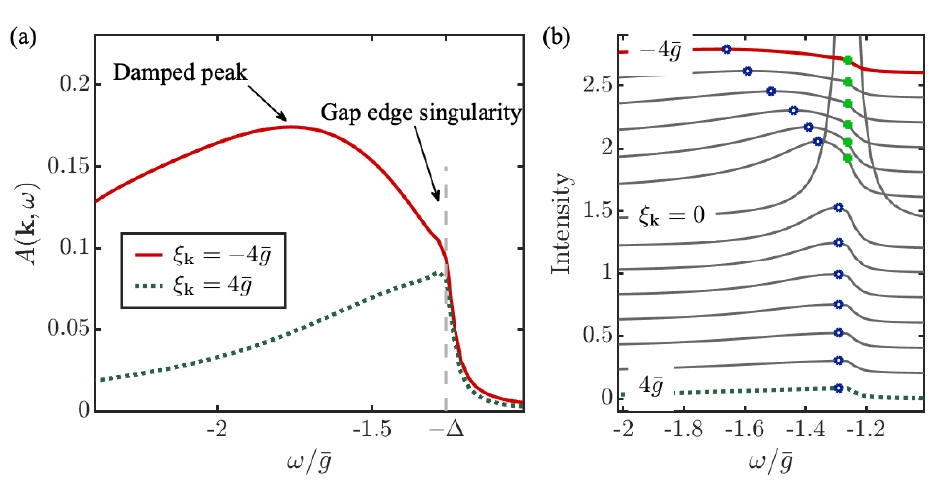}
\caption{(a) Spectral function  $A({\bm k}, \omega)$
at positive and negative  $\xi_{\bm k}=\pm 4{\bar g}$
at $\gamma=0.6$.
To account for impurity scattering, we convoluted the spectral function with a Lorentzian of width $\sim 0.03 {\bar g}$.
(b) Spectral function at a set of discrete momenta.
It 
displays a non-dispersing gap edge singularity (green dots) and a dispersing maximum (blue circles).
This theoretical $A({\bm k}, \omega)$ is consistent with the ARPES data for Bi2201, Ref.~\cite{he2011single} (see text).}
\label{fig:spectral_sc_2}
\end{figure}

The spectral function $A({\bm k}, \omega)$ is shown in Fig. (\ref{fig:spectral_sc}).
For comparison with ARPES, we set $\omega$ to be negative: $\omega = -(\Delta + \delta)$.
For any $\gamma$, there is no frequency range, where $A({\bm k}, \omega)$ is a $\delta$-function, simply because the bosonic mass vanishes at a QCP.
Still,  for $\gamma <1/2$, $D(-(\Delta + \delta))-1 \propto \delta$ and $Z(-(\Delta + \delta)) \approx Z(-\Delta) = Z(\Delta)$.
In this situation, the spectral weight on the Fermi surface, integrated over an infinitesimally small range
around $\omega =-\Delta$ immediately above the real axis, is finite, like in BCS case. Away from the Fermi surface,
the spectral function vanishes as $|\omega+\Delta|^{1/2-\gamma}$ at the gap edge and
displays a quasiparticle peak at $\omega \approx -(\Delta^2 + ({\xi_{\bm k}}/Z(\Delta))^2)^{1/2}$.
The peak is well defined at small $\delta$ as its width  $O(\delta^{1/2 -\gamma})$ is parametrically smaller than its frequency.
This is the same behavior as in
Fig.~\ref{fig:sche} (a).
For $\gamma >1/2$,
the
situation is qualitatively different. Now
$Z(-\Delta - \delta)$
diverges at $\delta \to 0$ and $D(-\Delta -\delta) -1 \sim |\delta|^\nu \ll |\delta|$. In this case, the integral of
$A({\bm k}_F, \omega)$ over an infinitesimally small range around $\omega = -\Delta$ vanishes, which can be interpreted as a vanishing of a quasiparticle peak.
At finite $\xi_{\bm k}$, the spectral function diverges at the gap edge as $1/|\omega +\Delta|^{\gamma/2 +\gamma -1}$.
For $\gamma$ slightly above $1/2$, $A({\bm k}, \omega)$ is non-monotonic and possess a broad maximum at
$|\omega + \Delta|  \sim \left( {{\xi_{\bm k}} / {\bar g}^\gamma} \right)^{1\over 1-\gamma}$.
This is the same behavior as in Fig.~\ref{fig:sche} (b). For larger $\gamma$, the maximum disappears, and  $A({\bm k}, \omega)$ monotonically decreases at $|\omega| > \Delta$.
This is the same behavior as in Fig.~\ref{fig:sche} (c).
For small $\xi_{\bm k}$, the maximum disappears at $\gamma \sim 0.9$.  For larger $\xi_{\bm k}$, it disappears
at smaller $\gamma$, first for positive $\xi_{{\bm k}}$ (see Fig.~\ref{fig:spectral_sc_2}). \\

{\it {\bf Comparison with ARPES}}~~~The behavior shown in Fig.~\ref{fig:spectral_sc_2} is our result in some range of $\gamma >1/2$.
For positive
$\xi_{\bm k}$
(i.e.,
outside the Fermi surface),
 the  spectral function has a single
non-dispersing maximum at the gap edge,
except for the smallest $\xi_{{\bm k}}$, while for negative
$\xi_{\bm k}$, $A({\bm k}, \omega)$ has a kink at the gap edge $\omega= -\Delta$ and a dispersing maximum at
$\omega = - \Delta - O\left(|{\xi_k}|^{1/(1-\gamma) }\right)$.
This behavior is  consistent with the ARPES data for Bi2201, Ref.~\cite{he2011single}.
The data shows that the spectral function near the antinode, where our analysis is valid, displays
an almost non-dispersing maximum at positive
$\xi_{\bm k}$, while for negative $\xi_{\bm k}$ it displays a non-dispersing kink at the same energy
and a dispersing maximum at larger $|\omega|$.
We associate the non-dispersing feature at both positive and negative $\xi_{\bm k}$ with the gap edge $\Delta$,
and associate the dispersing maximum, observed in~~\cite{he2011single} at $\xi_{\bm k} <0$, with the dispersing maximum in Fig.~\ref{fig:spectral_sc_2}.\\

{\it {\bf Discussion and summary.}}~~~
In this work, we analyzed the applicability of quasiparticle description of a superconducting state which emerges out of a non-Fermi liquid at a metallic QCP.
We considered the model with an effective dynamical 4-fermion interaction $V(\Omega) \propto 1/\Omega^\gamma$, mediated by a gapless boson at a QCP
and analyzed the spectral function and the DoS for $\gamma \in (0,1)$. Interaction $V(\Omega)$ gives rise to a non-Fermi liquid in the normal state with
self-energy $\Sigma (\omega) \propto \omega^{1-\gamma}$ and to pairing below some finite $T_c$.  A superconducting order gaps out low-energy excitations and, at a
first glance, should restore fermionic coherence.
We found, however, that this holds only for $\gamma <1/2$.
For larger $\gamma$ the spectral function and the DoS exhibit qualitatively different behavior than that in a superconductor with coherent quasiparticles.
(different power-laws).
We argued that the quasiparticle peak broadens up and completely disappears for $\gamma$ close to one.

Away from a QCP, a pairing boson is massive and at the lowest energies a  Fermi-liquid description holds already in the normal state and continue to hold in a superconductor. In particular, in the immediate vicinity of the gap edge, the system displays a BCS-like behavior for all $\gamma$.  Still, the system behavior over a broad frequency range is governed by the physics at a QCP,
as numerous experiments on the cuprates and other correlated systems indicate. We argued that our results are quite consistent with the ARPES data
for Bi2201~\cite{he2011single,hashimoto2014energy}.

\begin{acknowledgments}
We acknowledge with thanks useful conversations with a number of our colleagues.
This work was supported by the U.S. Department of Energy, Office of Science, Basic Energy Sciences, under Award No. DE-SC0014402.
\end{acknowledgments}

%

\clearpage

\widetext

\setcounter{figure}{0}
\renewcommand{\thefigure}{S\arabic{figure}}
\setcounter{equation}{0}
\renewcommand{\theequation}{S\arabic{equation}}

\begin{center}
\textbf{\large Supplementary information for ``Density of states and spectral function of a superconductor out of a quantum-critical metal''\\
by Shang-Shun Zhang and Andrey V Chubukov}
\end{center}

\section{Gap equation along the real-frequency axis and its solution}

We will use the approach pioneered by Marsiglio, Shossmann, and  Carbotte~\cite{Marsiglio_88sm}.
In this approach, one first solves non-linear gap equation along the Matsubara axis, which can be done rather
straightforwardly as the gap function $\Delta (\omega_m)$ can be chosen to be real for all frequencies and
is a regular function of $\omega_m$ even when the pairing boson is massless.
One then uses this $\Delta (\omega_m)$ as an input for the equation for complex $\Delta (\omega)$ along the real frequency axis.

\begin{figure}[!b]
\centering
\includegraphics[scale=1]{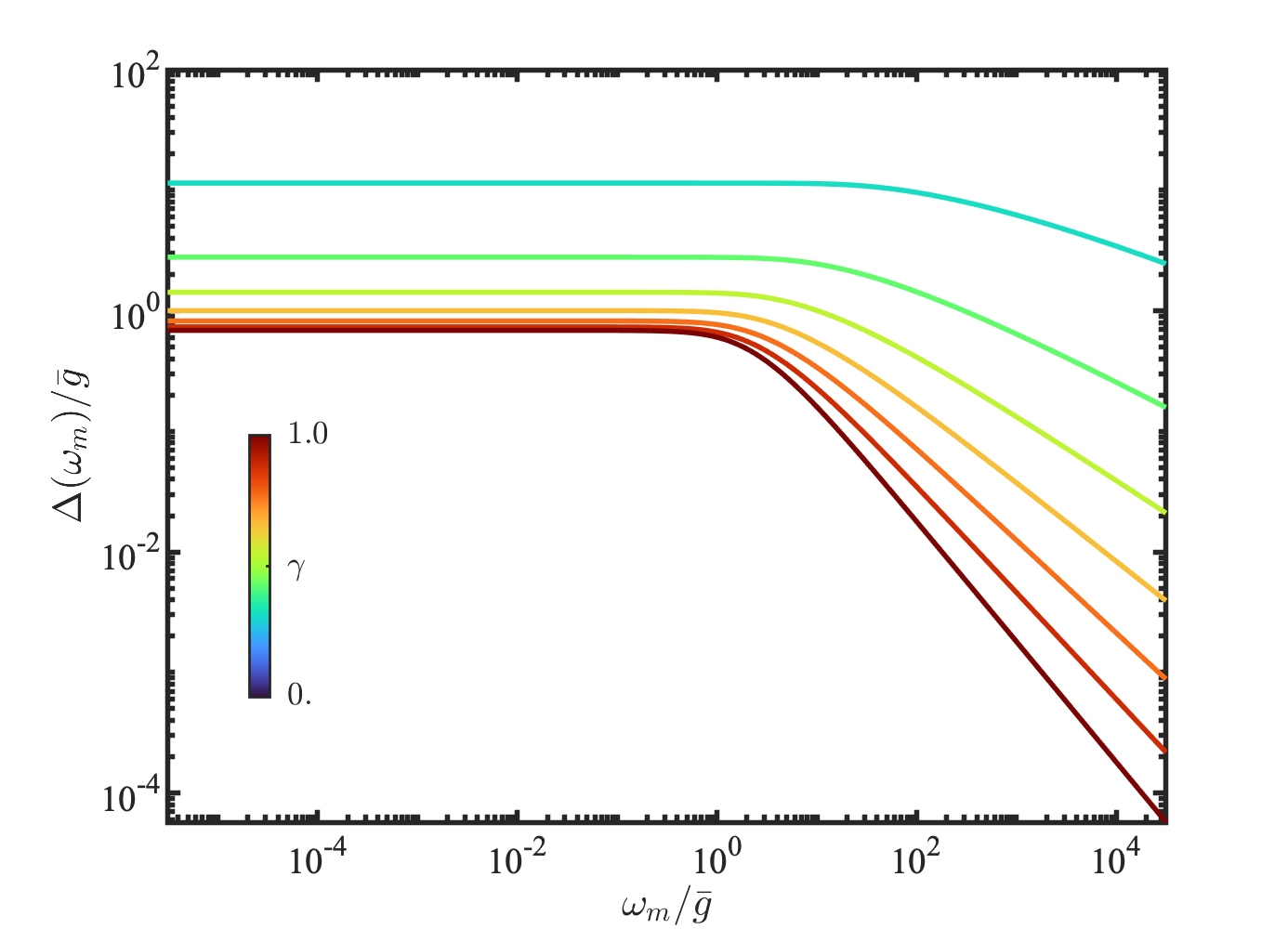}\caption{Numerical resulls for the gap function $\Delta(\omega_{m})$ along the Matsubara axis. The
calculation is performed at temperature $T=10^{-6}{\bar g}$ using the hybrid-frequency method
~\cite{paper_2}.
}
\label{fig:matsubara}
\end{figure}

The non-linear integral equation for $D(\omega_m) = \omega_m \Delta (\omega_m)$ on the Matsubara axis, and the equation  for
the inverse quasiparticle residue $Z(\omega_m) =1+\Sigma(\omega_m)/\omega_m$ ($\Sigma(\omega_m)$ is the fermionic self-energy), have the form
\beq
\omega_{m} D(\omega_{m})=\pi T \sum_{\omega_{m}^{\prime}} \frac{ \(D(\omega_{m}^{\prime})-D(\omega_{m})\) \text{sgn}(\omega_m^{\prime}) }{\sqrt{1 + D^{2}(\omega_{m}^{\prime}) } }V(\omega-\omega_m^{\prime}), \label{eq:gap_mats}
\eeq
\beq
Z(\omega_{m})= 1 + \frac{1}{\omega_m} \pi T \sum_{\omega_{m}^{\prime}} \frac{ \text{sgn} (\omega_m^{\prime} ) }{\sqrt{1+D^{2}(\omega_{m}^{\prime}) } } V(\omega-\omega_m^{\prime}) , \label{eq:Z_mats}
\eeq
where $V(\Omega_m) = \left( {\bar{g} }/{ \rvert \Omega_{m} \rvert }  \right)^{\gamma}$ is the same as in the main text ($\bar{g}$ is an effective fermion-boson coupling, and $\gamma$ depends on the underlying microscopic model).
This set of equations has a non-zero solution $D(\omega_m)$ below a finite pairing temperature $T_p \sim {\bar g}$.
Fig.~\ref{fig:matsubara} shows the numerical solution for $\Delta (\omega_m)$  at $T = 10^{-6}{\bar g} \ll T_p$, for different $\gamma$.  We see from the
figure
that $\Delta(\omega_m)$ approaches a  constant value  at small frequencies and decays as $\omega_m^{-\gamma}$ at high frequencies. This behavior holds for all $\gamma$ and can be easily verified analytically.

The gap equation along the real frequency axis is
\beq
\omega B(\omega) D(\omega) = A(\omega) + C(\omega),
\label{eq:gap_real1}
\eeq
(Eqn. (1) in the main text).
This equation is obtained by using the spectral representation
of an analytic function on the upper frequency half-plane
\beq
f (i\omega_m) =\frac{1}{\pi} \int dx  \frac{\text{Im} f(x)}{x - i \omega_m}
\eeq
and, where possible, keeping $D(\omega_m)$ as an input function.
This approach was pioneered for electron-phonon interaction in Refs.~\cite{Marsiglio_91,Karakozov_91,combescot,Wu_19_1}
for the electron-phonon problem.

For our case, the functions $A(\omega)$ and $B(\omega)$ are directly expressed via
$D(\omega_m)$ along the Matsubara axis as
\begin{align}
A(\omega) & =\frac{1}{2}\int_{0}^{\infty}d\omega_{m}\frac{D(\omega_{m})}{\sqrt{1+D^{2}(\omega_{m})}}\nonumber \\
 & \times\left(\frac{\bar{g}^{\gamma}}{(\omega_{m}+i\omega)^{\gamma}}+\frac{\bar{g}^{\gamma}}{(\omega_{m}-i\omega)^{\gamma}}\right),\label{eq:Aw}\\
B(\omega) & =1+\frac{i}{2\omega}\int_{0}^{\infty}d\omega_{m}\frac{1}{\sqrt{1+D^{2}(\omega_{m})}}\nonumber \\
 & \times\left(\frac{\bar{g}^{\gamma}}{(\omega_{m}+i\omega)^{\gamma}}-\frac{\bar{g}^{\gamma}}{(\omega_{m}-i\omega)^{\gamma}}\right).\label{eq:Bw}
\end{align}
 and $C(\omega)$ is  given by
 \begin{equation}
C(\omega)=\bar{g}^{\gamma}\sin\frac{\pi\gamma}{2}\int_{0}^{\omega}\frac{d\Omega}{\Omega^{\gamma}}
\frac{D(\omega-\Omega)-D(\omega)}{\sqrt{D^{2}(\omega-\Omega)-1}},\label{eq:Cw}
\end{equation}
 (Eqn (3) in the main text).
This function depends on the running $D(\omega -\Omega)$, which
 makes Eq. (\ref{eq:gap_real1}) an integral equation.
 The inverse residue $Z(\omega)$ is expressed via  $D(\omega')$ as
\bea
Z(\omega)= B(\omega) + \frac{\bar{g}^{\gamma}\sin\frac{\pi\gamma}{2}}{\omega}\int_{0}^{\omega}\frac{d\Omega}{\Omega^{\gamma}}
\frac{1}{\sqrt{D^{2}(\omega-\Omega)-1}}
\label{eq:gap_real2}
\eea
(Eqn (4) in the main text) and is readily obtained once $D(\omega)$ is known.

\begin{figure}
\centering
\includegraphics[scale=1]{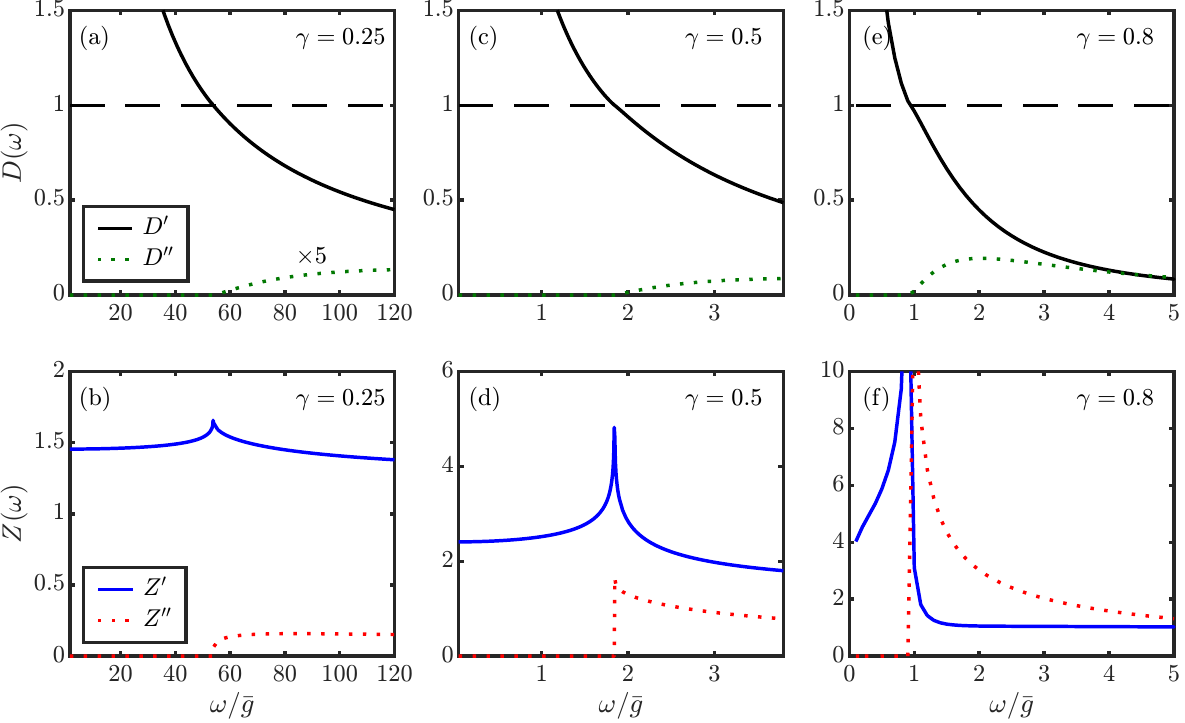}\caption{Numerical results for $D(\omega)$ and $Z(\omega)$ for $\gamma=0.25$, $\gamma=0.5$ and $\gamma=0.8$. }
\label{fig:solution}
\end{figure}

The gap equation along the real-frequency axis has an iterative structure in the sense that $D(\omega)$ depends on $D(\omega')$ at $\omega' < \omega$.
This  allows us to solve this equation iteratively, using the low-frequency form $D(\omega) \simeq \Delta(0)/\omega$ as an input, with $\Delta(0) \equiv  \Delta(\omega_m = \pi T)$.
In Fig.~\ref{fig:solution} we show the results for $D(\omega)$ and $Z(\omega)$ for three representative values of $\gamma$.
In all  cases, $D(\omega)$ and $Z(\omega)$ are real below the gap edge $\omega = \Delta$ and are complex above the gap edge,
where $\Delta$ is defined as $\Delta (\omega) =1$ at $\omega =\Delta$.
We see that for $\gamma >1/2$, $Z(\omega)$ diverges at the gap edge. We use this fact in the main text in the analysis of the spectral function.

\section{The case of $\gamma=1/2$}

In the main text we argued that for $\gamma <1/2$,  the function $D(\Delta -\delta)- 1 \propto \delta$, where
$\delta = \Delta - \omega$,
and the correction scales as $\delta^{3/2-\gamma}$.
More specifically, we found iteratively that
\beq
D(\Delta -\delta)  = 1 + \delta \sum_{n=0}^\infty \alpha_n \delta^{n \epsilon}
\label{eq:series1}
\eeq
where $\epsilon = 1/2 - \gamma$ and $\alpha_0 = O(1/{\bar g})$.
   The expression for $\alpha_1$ is presented in the main text, after Eq. (4).
   It is proportional to $J(\gamma,1) = B(1-\gamma, \gamma -3/2) - B(1-\gamma, \gamma -1/2)$, where
    $B(a,b)$ is a Beta function ($B(a,b) = \Gamma(a) \Gamma(b)/\Gamma(a+b)$).   For small $\epsilon$ (i.e., for $\gamma \leq 1/2$), $\alpha_1 \sim J(\gamma,1) \sim 1/\epsilon$.  For the next term in (\ref{eq:series1}) we find $\alpha_2 \sim 1/\epsilon^2$, and so on.

 We see that the perturbative expansion in $\delta^\epsilon$ in (\ref{eq:series1}) holds for
  $(\delta/{\bar g})^\epsilon/\epsilon \leq 1$.
Outside this range, all terms in Eq. (\ref{eq:series1}) are relevant.  As $\gamma$ approaches  $1/2$ from below and $\epsilon$ decreases,   the perturbative regime shrinks to exponentially small
$\delta < {\bar g} \exp(-|\log{\epsilon}|/\epsilon)$.

 To understand the form of $D(\omega)$ outside the perturbative regime, we express $(\delta/{\bar g})^\epsilon$ as $e^{\epsilon \log{(\delta/{\bar g})}}$ and expand
 (\ref{ss}) in powers of $\log{(\delta/{\bar g})}$.  We obtain
 \bea
D(\Delta - \delta) = 1 + \delta \sum_{n=0}^{\infty} {\tilde \alpha}_{n} (\log{\frac{\delta}{{\bar g}}})^n
\label{ss}
\eea
 where ${\tilde \alpha}_0 = \alpha_0 + \alpha_1 + \alpha_2+ ..$, ${\tilde \alpha}_1 = \epsilon \alpha_1 +
 2 \epsilon \alpha_2+ ..$,  ${\tilde \alpha}_2 = 2 \epsilon^2 \alpha_2 + ...$.   We see that each ${\tilde \alpha}_n$ is a  series, in which the first term is independent on $\epsilon$, and the others diverge as powers of $1/\epsilon$,
  because $\alpha_1 \sim 1/\epsilon$, $\alpha_2 \sim 1/\epsilon^2$, and so on.

  We now argue that  singular parts of ${\tilde \alpha}_n$ can be neglected.  The argument is two-fold.  First, in the calculations, the $1/\epsilon$  divergencies originate from the  divergence of  $J(1/2-\epsilon,1) \approx 1/(2\epsilon)$.  This divergence is regularized by a finite boson mass, such that strictly at $\epsilon =0$, one has $J(1/2, 1) =0$ instead of infinity.  Second, if we assume that ${\tilde \alpha}_0$ in (\ref{ss}) remains finite at $\epsilon =0$ and substitute the trial $D(\Delta - \delta) = 1 + \delta {\tilde \alpha}_0$ in the gap equation at $\gamma =1/2$ and  compute iteratively  the next term in $D(\Delta -\delta)$, we find it in the form
  ${\tilde \alpha}_1 \delta \log{(\delta/{\bar g})}$ with a finite
  $\tilde{\alpha}_1 = \sqrt{{\bar g}\tilde{\alpha}_0}/(4\Delta B(\Delta))$.   Extending the iterative analysis, we find that all ${\tilde \alpha}_n$ are finite at $\gamma =1/2$ (i.e., $\epsilon =0$),  as we anticipated.

\begin{figure}
\centering
\includegraphics[scale=1]{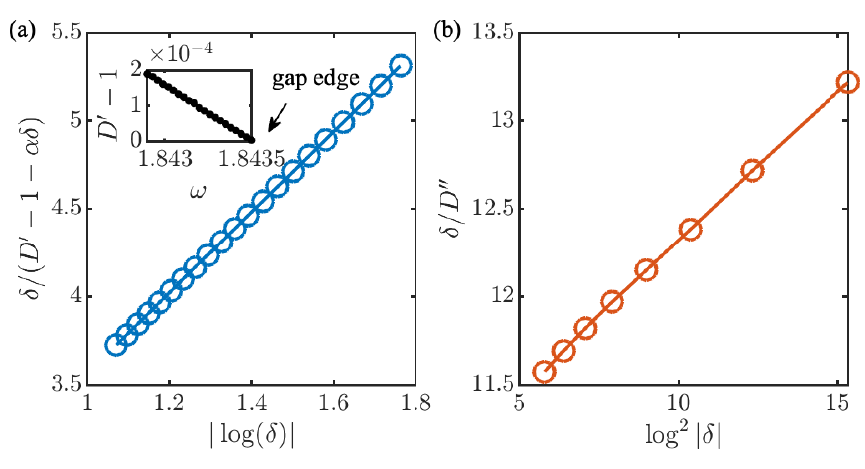}
\caption{The real and imaginary parts of the gap function near the gap edge $\omega=\Delta$ for $\gamma=1/2$, obtained by solving the non-linear gap equation numerically.
The leading term in $D' (\Delta - \delta) -1$, shown in
the inset of
(a),  is linear in $\delta  \Delta-\omega$, and the subleading
scales as $\delta/|\log{|\delta|}|$, 
as is confirmed by the linear relation in panel (a), 
The imaginary part
$D^{\prime\prime}$ appears 
at negative $\delta$ above the gap edge.
The numerical result
in panel (b) 
clearly shows the scaling relation
$D^{\prime\prime}\sim \delta/\log^2{(|\delta|/{\bar g})}$, expected from the Kramers-Kronig relation 
 with $D'$.}
\label{fig:1half_sol}
\end{figure}

We didn't manage to sum up analytically the logarithmic series in (\ref{ss}). The numerical solution for $D(\Delta -\delta)$ for $\gamma =1/2$ shows that $D(\Delta -\delta) -1$ remains linear in $\delta$
(see Fig.~\ref{fig:solution} (c)),
and the corrections scale as $1/|\log{(|\delta|/{\bar g})}|$ (see Fig.~\ref{fig:1half_sol} (a)).
By Kramers-Kronig relation, this implies that at negative $\delta$, when $\omega >\Delta$ is above the gap edge,
the imaginary part of $D(\omega)$ scales as $D^{''} (\Delta + |\delta|) \propto \delta/\log^2{(|\delta|/{\bar g})}$ (the same form is obtained by just noticing that
$\log({-|\delta|}) = \log({-(\omega -\Delta + i0)}) = \log{|\delta|} -i\pi$).
This form of  $D^{''} (\Delta + |\delta|)$ is consistent with our numerical solution above the gap edge, Fig.~\ref{fig:1half_sol} (b).
The solution clearly shows that the ratio $D^{''} (\Delta + |\delta|)/\delta$ decreases at the smallest $\delta$.

\begin{figure}
\centering
\includegraphics[scale=1]{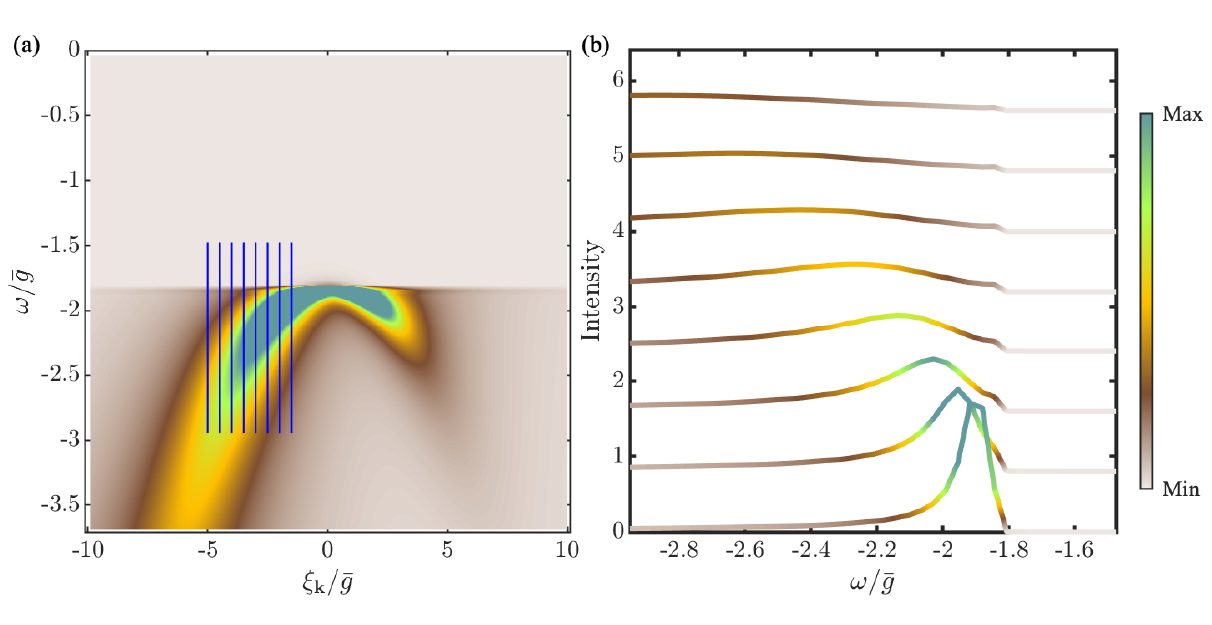}\caption{(a) The spectral function $A({\bm k}, \omega)$ for $\gamma=1/2$. (b) Constant $\xi_{\bm k}$ cuts along the blue lines in panel (a). }
\label{fig:1half}
\end{figure}

We next use the result for $D(\omega)$ to obtain the inverse quasi-particle residue near the gap edge.
Substituting $D(\Delta -\delta) \approx 1 + {\tilde \alpha_0} \delta$ into  (\ref{eq:gap_real2}), we obtain at $\gamma=1/2$
\bea \label{eq:Z2}
Z(\Delta - \delta) = {1\over 2\Delta} \sqrt{{{\bar g}\over \tilde{\alpha}_0 }}
|\log{\delta}|.
\eea
Analytically continuing this function to negative $\delta$, i.e., to $\omega$ above the threshold,
 we obtain
\begin{align}
\label{eq:Z12}
Z(\Delta + |\delta|)={1\over 2\Delta} \sqrt{{{\bar g}\over \tilde{\alpha}_0 }} \left( |\log{|\delta|}|
 + i \pi \right),
\end{align}
Note that the imaginary part of $Z(\omega)$ jumps to a finite value at $\omega$ infinitesimally above the threshold.
This behavior is consistent with the numerical solution for $Z(\omega)$, see Fig.~\ref{fig:solution} (d).

Finally, we use the results for $D(\omega)$ and $Z(\omega)$ and compute the spectral function
near the gap edge.
On the Fermi surface,  the spectral function at negative $\omega$ and $|\omega| > \Delta$ takes the form
\bea
A({\bm k}_F,\omega) \propto {1\over \rvert \omega + \Delta \rvert \log ({\bar g}/ \rvert \omega + \Delta \rvert )}.
\eea
Slightly away from the
Fermi surface,
the spectral function has a peak at $|\omega|  = \Delta+ \delta_{\bm k}$ where $\delta_{\bm k} \sim ( \xi^2_{\bm k} / {\bar g} )/ \log^2 (|{\bar g} / \xi_{\bm k}|)$.
The peak width  scales as $\delta_{\bm k} / \log (|{\bar g} / \xi_{\bm k}|)$ and is logarithmically smaller than the energy variation $|\omega| -\Delta$.
Also, for any non-zero $\xi_{\bm k}$, the spectral function jumps at the gap edge to a finite value  of order $1/\xi_{\bm k}^2$.
In Fig.~\ref{fig:1half} we show the numerical result for the spectral function. It is consistent with the
 behavior we just described.

\section{Universal form of the spectral function at $1/2<\gamma<1$}

\begin{figure}
\centering
\includegraphics[scale=1]{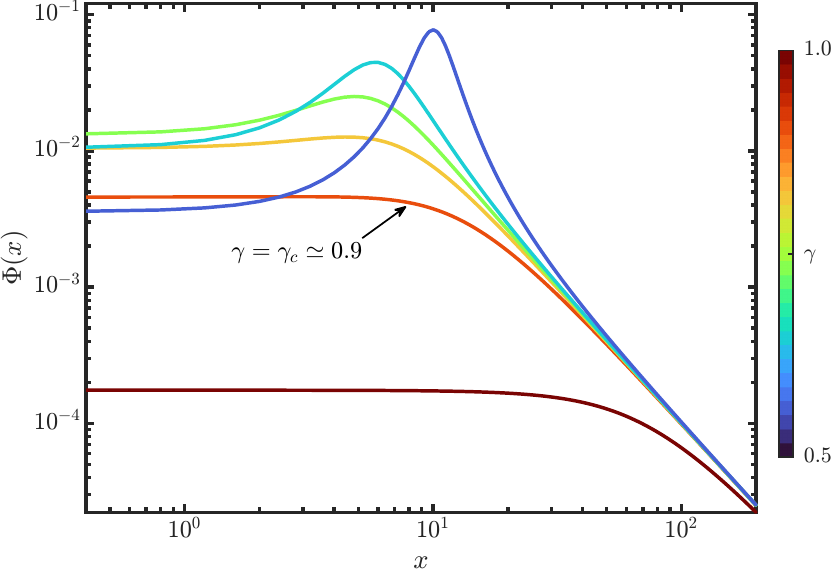}\caption{The function $\Phi(x)$, Eqn (\ref{ww}), for different values of $\gamma$. }
\label{fig:phi}
\end{figure}

For frequencies $\omega$ near the gap edge and for momenta near the Fermi surface, when $\xi_{\bf k}$ is much smaller than
$|\omega Z (\omega)|$, a  straightforward calculation shows that for $\gamma > 1/2$, the spectral function
can be expressed as a scaling function of $\xi_{\bf k}/|\omega + \Delta|^{1-\gamma} {\bar g}^\gamma$ (we set $\omega <0$).  Namely,
\begin{equation}
A({\bm k},\omega) \propto \frac{1}{\rvert \omega+\Delta \rvert^{\frac{\nu}{2}+1-\gamma}}\Phi\left(\frac{\xi_{\bm k}}{\rvert\omega+\Delta\rvert^{1-\gamma} {\bar g}^\gamma  }\right),
\end{equation}
where
\begin{equation}
\Phi(x)\equiv\frac{x^{2}+Q_\gamma^2\sin [\pi(\nu-c)] /\sin (\pi c) }{\left(x^{2}+Q_\gamma^2 \cos (2\pi\gamma) \right)^{2}+ Q_\gamma^4 \sin^{2} (2\pi\gamma) }
\label{ww}
\end{equation}
with $Q_{\gamma}=\sin(\pi\gamma/2)B(1-\gamma,\nu/2+\gamma-1)$.
In Fig.~\ref{fig:phi},
we plot the dimensionless function $\Phi(x)$ for different values of $\gamma$.
At $x \gg 1$, $\Phi(x) \sim 1/x^2$; at $x\ll 1$, $\Phi(x) \sim $ const.
For $\gamma < \gamma_c \simeq 0.9$, function $\Phi(x)$ contains a local maximum at
\bea
x_*^2 \sim \sqrt{(u-v)^2+w^2}-u,
\eea
where $u=Q_\gamma^2 \sin [\pi (\nu-c)]/\sin( \pi c)$, $v=Q_\gamma^2 \cos (2 \pi \gamma)$, and $w=Q_\gamma^2 \sin (2 \pi \gamma)$.
This maximum can be interpreted as
an over-damped, but still existing  quasi-particle peak.
At $\gamma > \gamma_c$, the function $\Phi(x)$ monotonically decreases with  $x$.
In this case, the  quasiparticle description breaks down completely.

We emphasize that this behavior holds only for small enough $\xi_{\bf k}$.
For larger $\xi_{\bf k}$, the spectral function does depend  on the sign of $\xi_{\bf k}$ and as $\gamma$ increases, the quasiparticle behavior gets completely destroyed first for positive $\xi_{\bf k}$ and then, at larger $\gamma$, for negative $\xi_{\bf k}$.

\end{document}